\documentclass[aps]{revtex4}

\usepackage{graphicx}
\usepackage{bm}
\usepackage{amssymb}
\usepackage{amsmath}
\usepackage{dcolumn}
\usepackage[latin1]{inputenc}
\usepackage[english]{babel}

\begin{document}


\title{Improvement of the Envelope Theory for Systems with Different Particles}

\author{Cyrille \surname{Chevalier}}
\email[E-mail: ]{cyrille.chevalier@umons.ac.be}
\thanks{ORCiD: 0000-0002-4509-4309}

\author{Cintia T. \surname{Willemyns}}
\email[E-mail: ]{cintia.willemyns@umons.ac.be}
\thanks{ORCiD: 0000-0001-8114-0061}

\author{Lorenzo \surname{Cimino}}
\email[E-mail: ]{lorenzo.cimino@umons.ac.be}
\thanks{ORCiD: 0000-0002-6286-0722}

\author{Claude \surname{Semay}}
\email[E-mail: ]{claude.semay@umons.ac.be}
\thanks{ORCiD: 0000-0001-6841-9850}

\affiliation{Service de Physique Nucl\'{e}aire et Subnucl\'{e}aire,
Universit\'{e} de Mons,
UMONS Research Institute for Complex Systems,
Place du Parc 20, 7000 Mons, Belgium}
\date{\today}

\begin{abstract}
\textbf{Abstract} The envelope theory is a method to compute approximate eigensolutions of quantum $N$-body Hamiltonians with a quite general structure in $D$ dimensions. The advantages of the method are that it is easy to implement and that $N$ is treated as any other parameters of the Hamiltonian, allowing the computation for systems of all sizes. If solutions are reliable, they are generally not very accurate. In the case of systems with identical particles for $D \ge 2$, it is possible to improve the precision of the eigenvalues by combining the envelope theory with a generalisation to $N$-body of the dominantly orbital state method. It is shown that a similar improvement can be achieved in the case of systems composed of identical particles plus a different one. The quality of the new procedure is tested with different systems.
\end{abstract}

\maketitle

\section{introduction}
\label{sec:intro}

Among all the methods to solve quantum many-body problems, the envelope theory (ET) \cite{hall80,hall83,hall04}, also known as the auxiliary field method \cite{silv10,silv12}, occupies a special place, because the computation cost is independent of the number of particles. It is quite easy to implement if all the particles are identical \cite{sema13} and becomes a little trickier if the number of different types of particles increases \cite{sema20}. Quite general type of Hamiltonians in $D$ dimensions can be considered \cite{sema17,sema18b,sema19} for which the ET can yield analytical bounds for the spectra in favourable situations. If this is not the case, numerical approximations with or without variational character can always be computed. 

The principle of the ET relies on the fact that the complete solution of a $N$-body harmonic oscillator Hamiltonian with two-body forces, say $H_\textrm{ho}$, can be obtained by the diagonalisation of a matrix of order $(N-1)$ \cite{hall79,silv10,cint01}. The idea is to build an auxiliary Hamiltonian, $\tilde H = H_\textrm{ho} + B$, where $B$ is a function of the masses and coupling constants of $H_\textrm{ho}$, whose structure is unequivocally determined from the general $N$-body Hamiltonian under study, say $H$. The approximate eigenvalues are obtained by an extremisation procedure performed on the eigenvalues of $\tilde H$ \cite{sema20}. These approximate values depend on characteristic global quantum numbers $Q$, whose the common structure given below implies a strong degeneracy of the levels, inherited from $H_\textrm{ho}$.  

In the case of systems composed of identical particles for $D\ge 2$, it is possible to partly correct this drawback and to improve the accuracy of the ET approximations by using two results stemming from the study of 2-body systems, but the price to pay is that the possible variational character of the approximate eigenvalues cannot longer be guaranteed. This has been done previously in two steps:
\begin{itemize}
\item In \cite{loba09}, it is shown that a universal effective quantum number exists for centrally symmetric 2-body systems. According to this result, a unique constant $\phi$ is introduced in the quantum number $Q$ of the ET (only one for identical particles) which yields improvements of the eigenvalues and observables, very significant for some Hamiltonians \cite{sema15a}. The optimal value of $\phi$ for the whole spectrum can be simply deduced from known accurate numerical results \cite{horn14}. 
\item In the dominantly orbital state method (DOSM), an  approximate solution is found by quantising the radial motion around a semiclassical solution for a purely orbital motion. Developed at the origin for 2-body systems \cite{olss97}, it has been extended to 3-body systems \cite{sema13b}. In \cite{sema15b}, the DOSM is generalised to $N$-body systems and its approximations are compared with the solutions from the ET to predict the value of $\phi$. Results are obtained in perfect agreement with those obtained in \cite{sema15a}. 
\end{itemize}

The purpose of this paper is to generalise these procedures for systems containing $N$ identical particles plus a different one, in order to extend the use of the ET for new systems in different physical contexts. For instance, in the Large-$N$ approach of QCD, the SU(3) gauge group is replaced by SU($N$), leading up to baryons made of $N$ quarks \cite{witt79}. To treat hadronic systems within this approach in combination with the constituent model \cite{will16} seems interesting in this energy domain where the usual expansion methods of the quantum field theory are not relevant. The use of the ET allows for a systematic study of baryons as a function of $N$ without the need of an enormous numerical cost, and the accuracy expected is sufficient for this kind of systems, as it appears in \cite{buis11,buis12}. So, an extension of the improved ET to $N+1$ systems will allow for computations for baryons containing a quark different from the others or for hybrid baryons containing identical quarks plus a constituent gluon. But, this technique could also be used in other domains where it is necessary to treat systems with a great number of particles. For instance, it could be used to estimate the binding energies of nuclei or clusters of cold atoms for which ab-initio calculations are already available, as for instance in \cite{gatt13,gatt11,kiev20}. The method has also allowed for the study of a possible quasi Kepler's third law for quantum many-body systems \cite{sema21}. 

The paper is organised as follows. The improvement of the ET with the DOSM for $N$ identical particles is recalled in Sec.~\ref{sec:Nid}. This section is developed to describe the principle of the calculations in a simpler context, but also to present new formulas. Indeed, the results given in \cite{sema15b} are correct but uselessly complicated, due to an unnoticed simplification. In Sec.~\ref{sec:Np1}, the improvement for the ET is extended to systems with all particles identical except one. The practical formulas to use are described and two coherence tests are performed. Different systems are studied in Sec.~\ref{sec:numtests} to show the relevance of this new method but also its limitations. Some concluding remarks are finally given in Sec.~\ref{sec:conclu} where some prospects are also presented. In the following, we work in natural units ($\hbar=c=1$) for $D \ge 2$. 

\section{$\bm N$ identical particles}
\label{sec:Nid}

For $N$ identical particles, the Hamiltonian considered is written
\begin{equation}
\label{HNid}
H=\sum_{i=1}^N T(|\bm{p}_i|) + \sum_{i<j=2}^N V(|\bm{r}_i-\bm{r}_j|),
\end{equation}
where $T$ is a kinetic energy (with some constraints \cite{sema18a}) and $V$ a two-body central potential, and where the centre of mass motion is removed ($\sum_{i=1}^N \bm p_i = \bm 0$). An ET approximate eigenvalue $E$ is given by the following compact set of equations for a completely (anti)symmetrised state of the auxiliary Hamiltonian $\tilde H$ associated with (\ref{HNid}):
\begin{subequations}
\label{ETNid}
\begin{align}
\label{ETNid1}
&E=N\, T(p_0) + C^2_N\, V (\rho_0), \\
\label{ETNid3}
&N\, T'(p_0)\, p_0 =  C^2_N\, V' (\rho_0)\, \rho_0, \\
\label{ETNid2}
&Q(N)=\sqrt{C^2_N}\,\rho_0\, p_0, 
\end{align}
\end{subequations}
where $U'(x)=dU/dx$, $C^2_N=N(N-1)/2$ is the number of pairs, $\rho_0$ and $p_0$ are the variables to be determined, and 
\begin{equation}
\label{QN}
Q(N) = \sum_{i=1}^{N-1} \left(2\, n_i + l_i + \frac{D}{2}\right) 
\end{equation}
is a global quantum number. This set is called compact because all the relevant variables ($\rho_0$ and $p_0$, see below) appear in three equations giving the expression for the energy (\ref{ETNid1}), the equation of motion (\ref{ETNid3}) and the rule for the quantisation (\ref{ETNid2}). The $\{n_i,l_i \}$ are the quantum numbers associated with the $N-1$ internal Jacobi coordinates \cite{silv10}. According to the nature of the particles, bosons or fermions, only some values are allowed for $Q(N)$. Their determination is a difficult problem, even for $N=3$ \cite{stan97}. But the ground states are determined for all values of $N$ in \cite{sema20}. Depending on the forms of $T$ and $V$, the approximate value $E$ can have a variational character \cite{sema20}. The method is easy to implement since it reduces to finding the pair ($\rho_0$,$p_0$) from a transcendental equation, and then computing $E$. Let us mention that \cite{sema15b}
\begin{equation}
\label{p0rho0}
p_0^2 = \frac{1}{N}\left\langle \sum_{i=1}^N \bm p_i^2 \right\rangle \quad \textrm{and}
\quad \rho_0^2 = \frac{1}{C^2_N}
\left\langle \sum_{i<j=2}^N (\bm r_i - \bm r_j)^2 \right\rangle, 
\end{equation}
where the mean value is computed with the eigenstate of the auxiliary Hamiltonian $\tilde H$ corresponding to $Q(N)$. So, $\rho_0$ can be interpreted as the mean distance between two particles and $p_0$ as the mean momentum of each particle. In \cite{sema13,sema20}, the variable $r_0= \sqrt{C^2_N} \, \rho_0$ is used because one-body potentials were also considered. It has been checked with the systems studied in \cite{sema15a} that the energies computed by (\ref{ETNid}) are the same that the ones computed with the extremisation procedure mentioned in the introduction. Nevertheless, the compact set~(\ref{ETNid}) is interesting because it yields directly the mean values $\rho_0$ and $p_0$ which are meaningful physical quantities. This set has a nice semiclassical interpretation \cite{sema13} and it is a compulsory step of the procedure explained below for the improvement by the DOSM:

\begin{enumerate}

\item In order to break the strong degeneracy inherent to the method, the global quantum number $Q(N)$ is shared into angular and radial contributions by the introduction of a parameter $\phi$
\begin{subequations}
\label{QNphi}
\begin{align}
\label{Qnulambda}
Q_\phi(N) &= \phi\, \nu + \lambda \quad \textrm{where} \\
\label{nulambda}
\nu &= \sum_{i=1}^{N-1} \left (n_i + \frac{1}{2} \right) \quad \textrm{and} \quad
\lambda =\sum_{i=1}^{N-1}\left (l_i + \frac{D-2}{2} \right).
\end{align}
\end{subequations}
This is inspired from the existence of an effective quantum number $q$ that determines with high accuracy the level ordering of centrally symmetric 2-body systems \cite{loba09}
\begin{equation}
\label{qnq}
q=\phi( n +1/2) +l+ (D -2)/2.
\end{equation}
The parameter $\phi$ in (\ref{qnq}) depends on the system, but not on quantum numbers $n$ and $l$ (for practical purposes, our definition of $\phi$ differs from the one used in \cite{loba09}). In (\ref{QNphi}), the same assumption is extended to $N\ge 2$. Note that the original quantum number $Q(N)$ is recovered with $\phi = 2$.

\item The second step is to obtain the DOSM solution of $H$. It has been shown in \cite{sema15b} that it can be computed from the ET compact set of equations. The pair ($\tilde \rho_0$,$\tilde p_0$) and the corresponding energy $\tilde E$ are computed for a purely collective orbital motion by solving (\ref{ETNid}) but with $Q(N)$ replaced by $\lambda$: 
\begin{subequations}
\label{ETNidtilde}
\begin{align}
\label{ETNid1tilde}
&\tilde E=N\, T(\tilde p_0) + C^2_N\, V (\tilde \rho_0), \\
\label{ETNid3tilde}
&N\, T'(\tilde p_0)\, \tilde p_0 =  C^2_N\, V' (\tilde\rho_0)\, \tilde\rho_0, \\
\label{ETNid2tilde}
&\lambda=\sqrt{C^2_N}\,\tilde\rho_0\, \tilde p_0.
\end{align}
\end{subequations}
Under the hypothesis that $\lambda \gg \nu$, a collective radial motion of the $N$ particles is introduced as a small perturbation by computing the energy from (\ref{ETNid1tilde}) with the following replacements  
\begin{equation}
\label{p0rho0star}
\tilde \rho_0 \to \tilde \rho_0+\Delta \rho \quad \textrm{and} \quad 
\tilde p_0 \to \sqrt{
p_r^2+
\frac{\lambda^2}{C^2_N (\tilde \rho_0+\Delta \rho)^2}
},
\end{equation}
inspired by semi-classical considerations developed in \cite{olss97}. With the assumptions given above, $\Delta \rho \ll \tilde \rho_0$ and $p_r \ll \tilde p_0 = \lambda/(\sqrt{C_N^2} \tilde \rho_0)$, where $\Delta \rho$ is a radial displacement and $p_r$ is a radial momentum. So, a power expansion of the radial contribution $\Delta E$ to $\tilde E$, limited to the lowest non-vanishing orders, gives
\begin{align}
\label{DeltaE}
\Delta E &\approx \frac{1}{2 \mu} p_r^2 + \frac{k}{2} \Delta \rho^2 \quad \textrm{with} \\
&\mu = \frac{\tilde p_0}{N\, T'(\tilde p_0)}, \nonumber \\
&k = \frac{2\, N\,\tilde p_0}{\tilde \rho_0^2}\,T'(\tilde p_0)
+ \frac{N\,\tilde p_0^2}{\tilde \rho_0^2}\,T''(\tilde p_0)
+  C^2_N\, V''(\tilde \rho_0).\nonumber 
\end{align}
The term in $\Delta \rho$ is canceled by the equation of motion~(\ref{ETNid3tilde}). This Hamiltonian must be quantised to compute the radial excitations. As $p_r$ and $\Delta \rho$
are effective collective variables, it is not clear if $\Delta \rho$, or a multiple of it, is the conjugate of $p_r$. Nevertheless, if we assume that $\Delta \rho$ is the conjugate variable of $p_r$, the solution is
\begin{equation}
\label{DeltaEq}
\Delta E \approx \sqrt{\frac{k}{\mu}}\, \left ( n + \frac{1}{2}\right).
\end{equation}
The effective quantum number $n$ must be reinterpreted in terms of the radial contribution $\nu$. The application of the formulas above to the case of the $N$-body harmonic oscillator, for which the DOSM gives the exact solution, shows that 
\begin{equation}
\label{ninterp}
\sqrt{C^2_N}\,\left ( n + \frac{1}{2}\right) = \nu .
\end{equation}
This constraint to recover the exact solution for the $N$-body harmonic oscillator allows to fix the problem about the identification of the correct pair of conjugate variables. Indeed, the same result for $\Delta E$ is obtained with another choice of the conjugate variables, but (\ref{ninterp}) will be different as in \cite{sema15b}. The complete solution for the DOSM is then $\tilde E + \Delta E$.

\item The system (\ref{ETNid}) is then solved with $Q(N) \to Q_\phi(N)= \lambda(1+\epsilon)$ where $\epsilon = \phi\,\nu/\lambda \ll 1$, in order to be in the same situation as the one of the previous item. If $\epsilon = 0$, the purely orbital solution $\tilde E$ is recovered. Starting from this solution, the energy is expanded in power of $\epsilon$, giving to the first order,
\begin{equation}
\label{Eeps}
\Delta E \approx N \, \tilde p_0 \,T'(\tilde p_0) \, \epsilon.
\end{equation}
This expression is identical to the one found in \cite{sema15b} but its form is simpler due to previously unnoticed vanishing contributions. The comparison of (\ref{DeltaEq}) and (\ref{ninterp}) with (\ref{Eeps}) gives directly the value of $\phi$
\begin{equation}
\label{phiid}
\phi = \frac{\lambda}{N\, \tilde{p_0}\, T'(\tilde{p_0})}\sqrt{\frac{k}{C_N^2\, \mu}},
\end{equation}
with $\mu$ and $k$ given by (\ref{DeltaE}). This value has been determined in the specific conditions where the radial motion has a small contribution to the energy with respect to the orbital motion, because it is computable with the DOSM. But since the structure of $\phi$ is assumed to be valid for any quantum numbers, (\ref{phiid}) can be used in principle for any state considered.

\item With the values $\lambda$ and $\nu$ chosen, and the value of $\phi$ computed with (\ref{phiid}), the system (\ref{ETNid}) is now solved with $Q(N)$ replaced by $Q_\phi(N)$ to yield the improved ET (IET) approximation. The original ET, with its possible variational solutions, is then recovered with $\phi=2$. For other values of $\phi$, the possible variational character of the solution cannot be guaranteed.

\end{enumerate}

As an example of Hamiltonian, let us consider the generic case 
\begin{equation}
\label{FGalphabeta} 
T(x)= F\, x^\alpha \quad \textrm{and} \quad V(x)=\mathrm{sgn}(\beta)\, G\, x^\beta,
\end{equation}
where $F$ and $\alpha$ are strictly positive numbers to satisfy the constraints in \cite{sema18a}, and where $\beta \ne 0$ and $G$ is a strictly positive number because of the sign function in $V$. The nonrelativistic case is covered with $F=1/(2m)$ and $\alpha=2$, and the ultrarelativistic one with $F=1$ and $\alpha=1$. The application of the above procedure gives 
\begin{equation}
\label{TVpower}
E=\mathrm{sgn}(\beta)\, (\alpha + \beta) \left( \left(\frac{C_N^2\,G}{\alpha}\right)^\alpha \left(\frac{N\,F}{|\beta|}\right)^\beta \left(\frac{Q_\phi(N)}{\sqrt{C_N^2}}\right)^{\alpha\,\beta}\right)^{1/(\alpha+\beta)}
\end{equation}
with the remarkably simple value
\begin{equation}
\label{phipower} 
\phi= \sqrt{\alpha+\beta}, 
\end{equation}
independent of $F$, $G$, $N$ and the global quantum number. According to the meaning of $T$, $E$ can be the mass of the system and then must be a positive number, or $E$ can be the binding energy and then can be negative or positive (for a confining potential). In this last case, bound states exist if $E$ and $\beta$ have the same sign. So, (\ref{TVpower}) implies that $\alpha+\beta > 0$, which is coherent with~(\ref{phipower}). If $E$ is the mass (as in the ultrarelativistic case), a strong attractive potential between the particles does not sound physically relevant and it seems natural to impose $\beta > 0$, and then $\alpha+\beta > 0$. When $\alpha=\beta=2$, the exact result for the $N$-body harmonic oscillator is recovered, as expected. The quality of (\ref{TVpower})-(\ref{phipower}) is tested for $N=3$ in Table~\ref{tab:plid}. The ``exact" results from \cite{basd90} are obtained with an elaborate hyperspherical expansion up to a grand orbital momentum $L=8$, which insures a good convergence of the expansion and a high accuracy of the eigenvalues \cite{rich81}.
\begin{table}[htb]
\protect\caption{Bosonic ground state energies for the case~(\ref{FGalphabeta}) with $N=3$, $\alpha=2$, $F=G=0.5$ and several values of $\beta$ (arbitrary units). The results from the ET (upper bound for $\beta < 2$ and lower bound for $\beta > 2$) and the IET with (\ref{phipower}) are compared with the ``exact" ones \cite{basd90}. The relative errors in \% are indicated between square brackets. In the case of $\beta=2$, ET and IET give the exact result as expected.}
\label{tab:plid}
    \begin{tabular}{ccrr}
        \hline
        $\beta$ & ``exact" & ET & IET \\
        \hline
        $-1$ & $-0.26675$ & $-0.12500\, [53]$ & $-0.28125\, [5.4]$ \\
        $-0.5$ & $-0.59173$ & $-0.49139\, [17]$ & $-0.59977\, [1.4]$ \\
        0.1 & 1.88019 & 1.91406\, [1.8] & 1.87743\, [0.15] \\
        0.5 & 2.91654 & 3.08203\, [5.7] & 2.90211\, [0.49] \\
         1 & 3.86309 & 4.08852\, [5.8] & 3.84130\, [0.56] \\
         2 & 5.19615 & 5.19615\, [0] & 5.19615\, [0] \\
         3 & 6.15591 & 5.68394\, [7.7] & 6.22479\, [1.1] \\
        \hline
    \end{tabular} 
\end{table}

The IET is tested on various systems in  \cite{sema15b} with a dramatic improvement in some cases. That is the reason why it is interesting to generalise the combination of the ET with the DOSM to systems with different particles. This is done in the next section.

\section{$\bm{N_a+1}$ systems}
\label{sec:Np1} 

The ET for Hamiltonians with different particles is detailed in \cite{sema20} and tested with systems containing $N_a$ particles of type $a$ and only one different particle of type $b$. The general Hamiltonian for such systems is written
\begin{equation}
\label{HNp1}
H=\sum_{i=1}^{N_a} T_a(|\bm{p}_i|) + T_b(|\bm{p}_N|) + \sum_{i<j=2}^{N_a} V_{aa}(|\bm{r}_i-\bm{r}_j|) + \sum_{i=1}^{N_a} V_{ab}(|\bm{r}_i-\bm{r}_N|),
\end{equation}
with $N=N_a+1$. The corresponding compact set of five ET equations, for an eigenstate completely (anti)symmetrised for the $N_a$ particles, is given in \cite{cimi21} 
\begin{subequations}
\label{Na1b}
\begin{align}
\label{Na1b1}
&E= N_a\, T_a(p_a') + T_b(P_0) + C_{N_a}^2\, V_{aa}(r_{aa}) + N_a\, V_{ab}(r_0'), \\
\label{Na1b2}
&N_a\, T_a' (p_a')\, \frac{p_a^2}{p_a'}= C_{N_a}^2 \,V_{aa}'(r_{aa})\, r_{aa} + \frac{N_a-1}{2} V'_{ab}(r_0') \frac{r_{aa}^2}{r_0'}, \\
\label{Na1b3}
&\frac{1}{N_a}\,  T'_a(p_a')\, \frac{P_0^2}{p_a'} + T_b' (P_0)\, P_0 = N_a\,  V'_{ab}(r_0')\, \frac{R_0^2}{r_0'},\\ 
\label{Na1b4}
&Q(N_a)=  \sqrt{C_{N_a}^2}\,  p_a\,  r_{aa},\\
\label{Na1b5}
&Q(2) = P_0 \, R_0,
\end{align}
with
\begin{equation}
\label{Na1b6}
p_a'^{\, 2}=p_a^2 + \frac{P^2_0}{N_a^2} \quad \textrm{and} \quad
r_0'^{\, 2}= R_0^2 + \frac{N_a-1}{2\,N_a}\,r_{aa}^2, 
\end{equation}
\end{subequations}
and with $Q(N)$ given by (\ref{QN}). $Q(N_a)$ is associated with the internal motions of the $N_a$ particles and its values are constrained by their bosonic or fermionic nature. But any value is allowed for $Q(2)$ which is associated with $R_0$, $R_0^2$ being the mean quadratic distance between the particle $b$ and the centre of mass of the $N_a$ particles \cite{sema20,cimi21}. Checks of (\ref{Na1b}) are provided in \cite{cimi21}.

To partially break the strong degeneracy implied by the global quantum numbers $Q(N_a)$ and $Q(2)$ in (\ref{Na1b}), and thus to improve the results of the ET, the four-step procedure described above for $N$ identical particles is generalised to the $N_a+1$ system in the four following subsections. 

\subsection{Separation between angular and radial dynamics}
\label{sec:angrad}

According to the situation for systems with all identical particles, the two global quantum numbers are splitted into an angular part and a radial part:
\begin{subequations}
\label{QNnulamb}
\begin{align}
\label{Qnulambdaa}
Q_\phi(N_a) &= \phi_a\, \nu_a + \lambda_a \quad \textrm{where} \\
\label{nla}
&\nu_a = \sum_{i=1}^{N_a-1} \left (n_i + \frac{1}{2} \right)\quad \textrm{and} \quad  
\lambda_a =\sum_{i=1}^{N_a-1}\left (l_i + \frac{D-2}{2} \right), \\
\label{Qnulambdab}
Q_\phi(2) &= \phi_b\, \nu_b + \lambda_b \quad \textrm{where} \\
\label{nla}
&\nu_b = n_b + \frac{1}{2} \quad \textrm{and} \quad  
\lambda_b =l_b + \frac{D-2}{2}. 
\end{align}
\end{subequations}
Two new parameters $\phi_a$ and $\phi_b$ are now introduced. With the nature of $Q(2)$ mentioned above, the quantum numbers $\{\nu_b,\lambda_b\}$ are linked to the relative motion between the particle $b$ and the centre of mass of the $N_a$ particles. The quantum numbers $\{\nu_a,\lambda_a\}$ are associated with a global excitation of the $N_a$ identical particles. 

\subsection{DOSM for $\bm{N_a+1}$ systems}
\label{sec:dosmNa}

The DOSM solution for a $N_a+1$ system is obtained in solving (\ref{Na1b}) with $Q(N_a)$ replaced by $\lambda_a$ and $Q(2)$ replaced by $\lambda_b$, that is to say in considering only orbital motions without radial excitations. An energy $\tilde E$ is  then computed with the associated variables $(\tilde p_a,\tilde r_{aa},\tilde P_0,\tilde R_0)$. Under the hypothesis that $\lambda_a \gg \nu_a$ and that $\lambda_b \gg \nu_b$, radial motions are introduced as small perturbations by computing the energy from (\ref{Na1b1}) with the following replacements  
\begin{equation}
\label{raapatilde}
\tilde r_{aa} \to \tilde r_{aa}+\Delta r, \quad  
\tilde p_a \to \sqrt{ p_r^2 + \frac{\lambda_a^2}{C^2_{N_a}(\tilde r_{aa}+\Delta r)^2} }
\end{equation}
and
\begin{equation}
\label{R0P0tilde}
\tilde R_0 \to \tilde R_0+\Delta R, \quad  
\tilde P_0 \to \sqrt{ P_r^2 + \frac{\lambda_b^2}{(\tilde R_0+\Delta R)^2} },
\end{equation}
in analogy with the replacement (\ref{p0rho0star}) made for all identical particles. With the assumptions given above, $\Delta r \ll \tilde r_{aa}$, $p_r \ll \tilde p_a$, $\Delta R \ll \tilde R_0$ and $P_r \ll \tilde P_0$. So, a power expansion of the sum $\Delta E$ of the radial contributions to $\tilde E$, limited to the lowest non-vanishing orders, gives after a lengthy calculation
\begin{align}
\Delta E &\approx \frac{1}{2}\left(\frac{1}{\mu_a} p_r^2 + \frac{1}{\mu_b} P_r^2 + k_a\, \Delta r^2 + k_b\, \Delta R^2 + k_c\, \Delta r \Delta R \right) \quad \textrm{with }\label{DEMSDOS}\\
& \mu_a = \frac{\tilde{p}_a'}{N_a T_a'(\tilde{p}_a')}, \notag \\
& \mu_b = \left( \frac{T_a'(\tilde{p}_a')}{N_a \tilde{p}_a'} + \frac{T_b'(\tilde{P}_0)}{\tilde{P}_0} \right)^{-1}, \notag \\
& k_a = \frac{N_a T''_a (\tilde{p}_a') \tilde{p}_a^4}{\tilde{r}_{aa}^2 \tilde{p}_a'^{\, 2}} + \frac{N_a T_a'(\tilde{p}_a')\tilde{p}_a^2}{\tilde{r}_{aa}^2} \left(\frac{3}{\tilde{p}_a'}-\frac{\tilde{p}_a^2}{\tilde{p}_a'^{\, 3}}\right) + C_{N_a}^2 V''_{aa}(\tilde{r}_{aa})\notag\\
& \text{\phantom{imm}}+ \frac{(N_a-1)^2 \tilde{r}_{aa}^2}{4 N_a \tilde{r}_0'^{\, 2}} V''_{ab}(\tilde{r}_0') + \frac{(N_a-1)}{2}\left( \frac{1}{\tilde{r}_0'} - \frac{(N_a-1)\tilde{r}_{aa}^2}{2 N_a \tilde{r}_0'^{\, 3}} \right)V_{ab}'(\tilde{r}_0'), \notag \\
& k_b = \frac{T''_a (\tilde{p}_a') \tilde{P}_0^4}{N_a^3 \tilde{R}_0^2 \tilde{p}_a'^{\, 2}} + \frac{T_b''(\tilde{P}_0)\tilde{P}_0^2}{\tilde{R}_0^2} + \frac{T_a'(\tilde{p}_a')\tilde{P}_0^2}{N_a \tilde{R}_0^2} \left(\frac{3}{\tilde{p}_a'} - \frac{\tilde{P}_0^2}{N_a^2 \tilde{p}_a'^{\, 3}}\right) \notag\\
& \text{\phantom{imm}} + \frac{2T'_b(\tilde{P}_0) \tilde{P}_0}{\tilde{R}_0^2} + \frac{N_a \tilde{R}_0^2}{\tilde{r}_0'^{\, 2}}V''_{ab}(\tilde{r}_0') + N_a\left(\frac{1}{\tilde{r}_0'}-\frac{\tilde{R}_0^2}{\tilde{r}_0'^{\, 3}}\right)V'_{ab}(\tilde{r}_0'), \notag\\
& k_c = \frac{2\tilde{p}_a^2 \tilde{P}_0^2}{N_a \tilde{p}_a'^{\, 2} \tilde{r}_{aa} \tilde{R}_0} \left( T_a''(\tilde{p}_a')- \frac{T_a'(\tilde{p}_a')}{\tilde{p}_a'}\right) + \frac{(N_a-1)\tilde{r}_{aa} \tilde{R}_0}{\tilde{r}_0'^{\, 2}} \left(V_{ab}'' (\tilde{r}_0') - \frac{V_{ab}'(\tilde{r}_0')}{\tilde{r}_0'} \right). \notag
\end{align}
There is no term in $\Delta r$ nor in $\Delta R$, but a term $\Delta r \Delta R$ couples the two radial motions. This energy $\Delta E$ must be quantised. From the nature of the variable $R_0$, it is clear that $\Delta R$ is the conjugate variable of $P_r$. For $\Delta r$ and $p_r$, the situation is similar to the case of all identical particles: $\Delta r$ is taken as the conjugate variable of $p_r$. The solution of the two one-dimensional coupled oscillators is given by (\ref{todco}) in Appendix~\ref{sec:tco}. In order to interpret correctly the two quantum numbers $n$ and $n'$ associated with (\ref{todco}), the formulas above are applied to the equivalent harmonic oscillator whose exact solution is given by formula~(37) in \cite{sema20} without one-body potentials. The comparison between the DOSM solution and the exact one shows that 
\begin{equation}
\label{ninterp2}
\sqrt{C^2_{N_a}}\,\left ( n + \frac{1}{2}\right) = \nu_a \quad \textrm{and} \quad
\left (n' + \frac{1}{2}\right) = \nu_b .
\end{equation}
The complete solution for the DOSM is then 
\begin{equation}
\label{spectrum_MEOD_Nap1_2}
E_{\textrm{DOSM}} = \tilde{E} + \sqrt{\frac{A}{C^2_{N_a} \mu}} \sum_{i=1}^{N_a-1} \left(n_i + \frac{1}{2}\right)+ \sqrt{\frac{B}{\mu}} \left(n_b + \frac{1}{2}\right), 
\end{equation}
with $\mu$, $A$ and $B$ given by (\ref{Etodco2}).

In order to check these formulas, we can consider the limit in which all particles are identical, that is to say $T_a(x)=T_b(x)=T(x)$ and $V_{aa}(x)=V_{ab}(x)=V(x)$. In this case, an eigenstate must be completely (anti)symmetrised. So, we must impose $r_{aa}=r_0'$ and $p_a'=P_0$, and as a consequence, $R_0^2 = (N_a+1)r_{aa}^2/(2 N_a)$ and $p_a^2 = (N_a^2-1)P_0^2/N_a^2$ \cite{cimi21}. These constraints must then be imposed on both orbital and radial motions. Indeed, it can be shown that (\ref{DEMSDOS}) reduces to (\ref{DeltaE}) when it is imposed that $\tilde{R}_0^2 = (N_a+1)\tilde{r}_{aa}^2/(2 N_a)$, $\Delta R^2 = (N_a+1)\Delta r^2/(2 N_a)$, $\tilde{p}_a^2 = (N_a^2-1)\tilde{P}_0^2/N_a^2$ and $p_r^2 = (N_a^2-1)P_r^2/N_a^2$. Let us note that this limit cannot be checked in (\ref{spectrum_MEOD_Nap1_2}), since no constraint has been imposed on the the radial motions controlled by $\Delta R$ and $\Delta r$ to obtain this formula.

\subsection{ET for $\bm{N_a+1}$ systems with small radial motions}
\label{sec:smallrad}

To compute the solutions of (\ref{Na1b}) for small radial excitations, $Q(N_a)$ is replaced by $Q_\phi(N_a) = \lambda_a(1+\epsilon_a)$ where $\epsilon_a = \phi_a\,\nu_a/\lambda_a \ll 1$, and $Q(2)$ by $Q_\phi(2) = \lambda_b(1+\epsilon_b)$ where $\epsilon_b = \phi_b\,\nu_b/\lambda_b \ll 1$. The case $\epsilon_a = \epsilon_b = 0$ corresponds to the purely orbital solution of Sec.~\ref{sec:dosmNa}. So, we can write $r_{aa} = \tilde r_{aa}+\Delta r$ with $\Delta r \ll \tilde r_{aa}$, $p_a =\tilde p_a +\Delta p$ with $\Delta p \ll \tilde p_a$, $R_0 = \tilde R_0+\Delta R$ with $\Delta R \ll \tilde R_0$ and $P_0 = \tilde P_0 + \Delta P$ with $\Delta P \ll \tilde P_0$. Equations~(\ref{Na1b4}) and (\ref{Na1b5}) give a link between $\Delta r$, $\Delta p$ and $\epsilon_a$, and between $\Delta R$, $\Delta P$ and $\epsilon_b$. Then, (\ref{Na1b1}) can be developed at first order in $\epsilon_a$ and $\epsilon_b$. All these lengthy calculations give
\begin{equation}
\label{Eepsab}
E \approx \tilde{E} + T'_a (\tilde{p}_a') \frac{N_a \tilde{p}_a^2}{\tilde{p}_a'} \epsilon_a + \left(T'_a (\tilde{p}_a') \frac{\tilde{P}_0^2}{N_a \tilde{p}_a'} + T_b'(\tilde{P}_0) \tilde{P}_0\right) \epsilon_{b}.
\end{equation}	
In order to check this formula, we again consider the limit in which all particles are identical as in the previous subsection. Considerations about the solution $(\tilde p_a,\tilde r_{aa},\tilde P_0,\tilde R_0)$ are unchanged, and we have to deal with the quantities $\epsilon_a$ and $\epsilon_b$. When all particles are identical, only the sums $Q(N_a)+Q(2)=Q(N_a+1)$ and $\lambda_a + \lambda_b$ are relevant, which implies that $\epsilon_a = \epsilon_b$. This is not surprising since these two numbers are ratios between radial and orbital excitations. Under these constraints, (\ref{Eepsab}) reduces to (\ref{Eeps}), as expected.

Making explicit $\epsilon_a$ and $\epsilon_b$ in terms of $\phi_a$ and $\phi_b$, (\ref{Eepsab}) is written
\begin{align}							  
&E \approx  \tilde{E} + D_a \frac{\phi_a}{\lambda_a} \sum_{i=1}^{N_a-1}\left(n_i+\frac{1}{2}\right) + D_b \frac{\phi_b}{\lambda_b}\left(n_b+\frac{1}{2}\right) 
\quad \textrm{with}
\label{Dev_TE_Na+1_fin}\\
&D_a = T'_a (\tilde{p}_a') \frac{N_a \tilde{p}_a^2}{\tilde{p}_a'}\quad \textrm{and} \quad D_b = T'_a (\tilde{p}_a') \frac{\tilde{P}_0^2}{N_a \tilde{p}_a'} + T_b'(\tilde{P}_0) \tilde{P}_0. \notag
\end{align}
The comparison of (\ref{spectrum_MEOD_Nap1_2}) with (\ref{Dev_TE_Na+1_fin}) gives
\begin{equation}
\label{phiaphib}
\phi_a = \frac{\lambda_a}{D_a} \sqrt{\frac{A}{C_{N_a}^2 \mu}} \quad \textrm{and} \quad \phi_b = \frac{\lambda_b}{D_b} \sqrt{\frac{B}{\mu}}.
\end{equation}
The quantities $\mu$, $A$, $B$, $D_a$ and $D_b$ can be computed from the values of the solution $(\tilde p_a,\tilde r_{aa},\tilde P_0,\tilde R_0)$ fixed by the choice of the collective orbital quantum numbers $\lambda_a$ and $\lambda_b$. If the complete expressions for $\phi_a$ and $\phi_b$ are quite complicated, their computation is straightforward.

\subsection{Summary of the procedure}
\label{sec:sumproc}

Once the values $\lambda_a$, $\lambda_b$, $\nu_a$ and $\nu_b$ are chosen, the value of $\phi_a$ and $\phi_b$ can be computed with (\ref{phiaphib}). The system (\ref{Na1b}) is then solved with $Q(N_a)$ replaced by $Q_\phi(N_a)$ and $Q(2)$ by $Q_\phi(2)$ given by (\ref{QNnulamb}) to yield the IET. If a solution of the original ET has a variational character, this cannot be guaranteed for the corresponding one obtained by the IET. 

By comparing the procedures described in Sec.~\ref{sec:Nid} and Sec.~\ref{sec:Np1}, one can appreciate the increase in complexity by simply adding one different particle to a set of identical ones. The system~(\ref{Na1b}) is so complex that no analytical solution was found, except in the case of harmonic oscillators. For this last system, it has been checked that the exact solution (formula~(37) in \cite{sema20} without one-body potentials) is recovered by this procedure and that $\phi_a =\phi_b =2$ in this case, as expected. 

\section{Numerical tests}
\label{sec:numtests}

For systems with identical particles, it was possible to test the ET ($D=1$ and 3) and the IET ($D=3$) \cite{sema15a,sema15b,sema19} thanks to the existence of very accurate results on simple many-body systems in the literature \cite{horn14,timo12,timo17}. Accurate results for simple systems with one particle different from all others are more scarce. In the following, the many-body systems considered in \cite{sema20} with $D=3$ are studied here with the ET and the IET. As no analytical solutions were found for the set~(\ref{Na1b}) for these systems, all calculations are performed numerically. 

\subsection{Three-body ultrarelativistic oscillators}
\label{sec:URoh}

The first system considered is composed of 3 ultrarelativistic bosons with a vanishing mass, interacting via harmonic oscillator potentials. The Hamiltonian in arbitrary units is written as
\begin{equation}
\label{Huoh}
H = \sum_{i=1}^{3} |\bm p_i| + (\bm r_1 -\bm r_2)^2 + \kappa \sum_{i=1}^{2} (\bm r_i -\bm r_3)^2.
\end{equation}
When $\kappa \ne 1$, the 3rd particle is different since it has a different interaction with the two other ones. Accurate numerical solutions for this Hamiltonian do not seem to be available in the literature. Fortunately, it is shown in \cite{salo20} that such solutions can be obtained from a rescaling of solutions for a nonrelativistic 3-body systems with a linear potential. This latter property can also be seen on the ET and IET solutions~(\ref{TVpower}) and (\ref{phipower}) where it is clear that the energy is invariant under the exchanges $\alpha \leftrightarrow \beta$ and $F \leftrightarrow G$ when $\beta >0$ and $N=3$. The accurate solutions of (\ref{Huoh}) presented in Table~\ref{tab:URoh} are properly rescaled nonrelativistic accurate results provided by J.M. Richard. He computed them with the previously mentioned method developed in \cite{rich81,basd90}. They are compared in Table~\ref{tab:URoh} with the upper bounds obtained with the ET and the results computed with the IET. For two values of $\kappa$, the three states presented have a positive parity with a vanishing total orbital angular momentum, the lowest one being the ground state. The quantum numbers $\nu_a$, $\lambda_a$, $\nu_b$ and $\lambda_b$ are assigned thanks to the knowledge of the global quantum numbers and the symmetry between the two identical particles of the exact solutions.

\begin{table}[htb]
\protect\caption{Positive parity spherical states of Hamiltonian~(\ref{Huoh}) for two values of $\kappa$. The results from the ET (upper bounds) and the IET with ($\phi_a,\phi_b$) and the global numbers $\langle \nu_a, \lambda_a, \nu_b, \lambda_b \rangle$ (see text) are compared with the ``exact" ones (see text). The relative errors in \% are indicated between square brackets.}
\label{tab:URoh}
\begin{tabular}{ccccccccccc}
\hline
\multicolumn{5}{c}{$\kappa=0.1$} & \ &  \multicolumn{5}{c}{$\kappa=10$} \\
\cline{1-5}\cline{7-11}
``exact" & ET & IET & $(\phi_a,\phi_b)$ & $\langle \nu_a, \lambda_a, \nu_b, \lambda_b \rangle$ & \ & ``exact" & ET & IET & $(\phi_a,\phi_b)$ & $\langle \nu_a, \lambda_a, \nu_b, \lambda_b \rangle$ \\
\hline
5.288 & 5.597\, [5.8] & 5.307\, [0.4] & (1.76,1.79) & $\langle 1/2,1/2,1/2,1/2 \rangle$ & & 14.506 & 15.353\, [5.8] & 14.699\, [1.3] & (1.82,1.80) & $\langle 1/2,1/2,1/2,1/2 \rangle$ \\
6.570 & 6.970\, [5.9] & 6.571\, [0.0] & (1.76,1.79) & $\langle 1/2,1/2,3/2,1/2 \rangle$ & & 19.134 & 20.272\, [5.9] & 19.291\, [0.8] & (1.82,1.80) & $\langle 3/2,1/2,1/2,1/2 \rangle$ \\
7.515 & 7.868\, [4.7] & 7.625\, [1.5] & (1.76,1.79) & $\langle 1/2,3/2,1/2,3/2 \rangle$ & & 20.340 & 21.580\, [6.1] & 21.032\, [3.4] & (1.82,1.80) & $\langle 1/2,3/2,1/2,3/2 \rangle$ \\
\hline
\end{tabular}
\end{table} 

The accuracy obtained by the ET upper bounds are around 4-6\%, while the relative error can be divided by more than 10 for the IET results. Even if its magnitude seems quite unpredictable, there is always an improvement with the IET. For both values of $\kappa$, the parameters $\phi_a$ and $\phi_b$ are close to $\sqrt{3}\approx 1.732$, the value (\ref{phipower}) for the case of identical particles ($\kappa = 1$). The values of $(\phi_a,\phi_b)$ depend on $\kappa$ but are independent of the quantum numbers of the states considered. This situation is similar to the one for all identical particles. Supplementary calculations would be desirable to see if this property is preserved for all the spectrum and for an arbitrary number of particles. Further comments about the accuracy of these results are given in the following section.

\subsection{Nonrelativistic three-body systems with a power law potential}
\label{sec:powlaw}

The second system is composed of three bosons interacting via different power law potentials, the 3rd one having a different mass. The Hamiltonian in arbitrary units is written as
\begin{equation}
\label{Hpl}
H = \sum_{i=1}^{2} \frac{\bm p_i^2}{2} + \frac{\bm p_3^2}{2 m}
+ \frac{1}{2}\, \textrm{sgn}(\beta) \sum_{i< j=2}^{3} \bm r_{ij}^\beta .
\end{equation}
Variational bounds from the ET and the results computed with the IET for the bosonic ground states are given in Table~\ref{tab:pl}. The very accurate (``exact") results are taken from \cite{basd90}, and have been computed with the previously mentioned method developed in \cite{rich81}.

\begin{table}[htb]
\protect\caption{Bosonic ground state energies of Hamiltonian~(\ref{Hpl}) for several values of $m$ and $\beta$. The results from the ET (upper bound for $\beta < 2$ and lower bound for $\beta > 2$) and the IET with ($\phi_a,\phi_b$) are compared with the ``exact" ones \cite{basd90}. The relative errors in \% are indicated between square brackets. In the case of $\beta=2$, ET and IET give the exact result as expected. Results for $m=1$ are given in Table~\ref{tab:plid}.}
\label{tab:pl}
    \begin{tabular}{cccrrcccrrcc}
        \hline
 & & \multicolumn{4}{c}{$m=0.2$} & &  \multicolumn{4}{c}{$m=5$} \\
\cline{3-6}\cline{8-11}
        $\beta$ & \ & ``exact" & ET & IET & $(\phi_a,\phi_b)$ & \ &  ``exact" & ET & IET & $(\phi_a,\phi_b)$\\
        \hline
        $-1$ & &$-0.1398$ & $-0.0645\, [54]$ & $-0.1316\, [5.9]$ & (1.07,1.14) &  \ & $-0.3848$ & $-0.1797\, [53]$ & $-0.3029\, [21]$ & (1.05,1.64)\\
        $0.1$ & & 1.9452 & 1.9804\, [1.8] & 1.9489\, [0.2] & (1.55,1.53) &  \ & 1.8486 & 1.8820\, [1.8] & 1.8568\, [0.4] & (1.48,1.77) \\
        $1$ & & 4.9392 & 5.2278\, [5.8] & 4.9687\, [0.6] & (1.79,1.77) &  \ & 3.4379 &  3.6386\, [5.8] & 3.4753\, [1.1] & (1.74,1.88)\\
        $2$ & & 7.5730 & 7.5730\, [0] & 7.5730\, [0] & (2,2) &  \ & 4.3729 &  4.3729\, [0] & 4.3729\, [0] & (2,2)\\
        $3$ & & 9.7389 & 8.9925\, [7.7] & 9.6703\, [0.7] & (2.16,2.20) & \ &  5.0166 & 4.6320\, [7.7] & 4.9693\, [0.9] & (2.20,2.15)\\
        \hline
    \end{tabular} 
\end{table}

The accuracy obtained by the ET bounds is better than 8\%, except for the Coulomb case. Except for one case, the Coulomb interaction with $m=5$, the improvement is again quite large with the IET, even if its magnitude seems quite unpredictable. The parameters $\phi_a$ and $\phi_b$ have a smooth variation with $\beta$ and have values close to $\sqrt{2+\beta}$, the value (\ref{phipower}) for all identical particles. It is a little bit frustrating to note that the results obtained here for the IET are slightly less good that for the naive choice $\phi_a=\phi_b=\sqrt{2+\beta}$ made in \cite{sema20}. But this trial had no a firm justification, contrary to the calculations developed here. 

Hamiltonians~(\ref{Huoh}) and (\ref{Hpl}) have both interacting parts of power law form. Results from Tables~\ref{tab:plid}, \ref{tab:URoh} and \ref{tab:pl} shows that very good results can be obtained when both exponent are positive. When the potential is singular, the accuracy is much lower. In Table~\ref{tab:plid}, the comparison between $\beta=-1$ and $\beta=-0.5$ indicates that the accuracy for systems with identical particles is deteriorated when the singularity increases. Nevertheless, the results are remarkably good in some situations. Further comments are given in Sec.~\ref{sec:conclu}. 

\subsection{Atoms}
\label{sec:atoms}

In atomic units, the Hamiltonian for $N_e$ electrons in an atom of charge $Z$ is written as
\begin{equation}
\label{HAt}
H = \frac{1}{2} \sum_{i=1}^{N_e} \bm p_i^2 + \frac{1}{2 m} \bm p_{N}^2 
- Z\sum_{i=1}^{N_e} \frac{1}{|\bm r_{iN}|} + \sum_{i< j=2}^{N_e} \frac{1}{|\bm r_{ij}|},
\end{equation}
where $N=N_e+1$ is the number of the nucleus with a mass $m$. Energies in eV are obtained by multiplying the eigenvalues of $H$ by the usual factor $\alpha^2\, m_e = 27.21$~eV. Hamiltonian~(\ref{HAt}) contains the main contributions to the binding energy in an atom. So, the approximate results are compared with the experimental data about ionisation energies \cite{nist} which are very close to the eigenvalues of (\ref{HAt}). Due to the mixing of attractive and repulsive parts, the approximate eigenvalues have no variational character. Taking into account the fermionic nature of the electrons (see Appendix~\ref{sec:gs}), ground state binding energies for some atoms computed with the ET and the IET are presented in Table~\ref{tab:At} with the experimental values. Note that the two electrons in $^4$He, $^6$Li$^+$, $^{12}$C$^{4+}$ and $^{16}$O$^{6+}$ can be treated as bosons because of the double-degeneracy due to the spin. 

\begin{table}[htb]
\protect\caption{Ground state binding energies (in eV) of Hamiltonian~(\ref{HAt}) for some atoms with two or more electrons. Results from the ET and the IET with ($\phi_a,\phi_b$) are compared with the experimental values (see text).}
\label{tab:At}
\begin{tabular}{rrrrc}
\hline
 & Exp. & ET & IET & $(\phi_a,\phi_b)$ \\
\cline{2-5}
$^4$He & 79 & 33 & 47 & $(1.21,1.78)$ \\
$^6$Li$^+$ & 198 & 85 & 123 & $(1.18,1.77)$ \\
$^6$Li &  203 & 66 & 95 & $(1.03,1.99)$ \\
$^{12}$C$^{4+}$ & 882 & 386 & 568 & $(1.16,1.77)$ \\
$^{12}$C & 1030 & 321 & 496 & $(0.96,2.10)$ \\
$^{16}$O$^{6+}$ & 1611 & 707 & 1047 & $(1.15,1.77)$ \\
$^{16}$O & 2044 & 672 & 1062 & $(0.94,2.14)$ \\
\hline
\end{tabular}
\end{table}

It is clear that the poor quality of the ET energies are not significantly improved by the IET. Curiously, the naive choice $\phi_a=\phi_b=1$ made in \cite{sema20} improves very much the binding energies but only when there is only two electrons in the atom. Even if it produces better results in some cases, this trial had no a firm justification, contrary to the calculations developed here. We identify three possible sources for this problem: the action of the Pauli principle; the mixing of attractive and repulsive parts in the potentials, forbidding the existence of a definite variational character for the solutions; the singularity of the Coulomb interaction. 

The problem could arise from the fermionic nature of the electrons, but results are not better when the two only electrons are treated as bosons. Moreover, good results are obtained for the fermionic ground state of the one-dimensional Calogero model with linear forces supplemented by inverse-cube forces \cite{sema19}. So, the origin of the problem should be found elsewhere. 

Some preliminary calculations with different three-body systems have been undertaken to clarify the situation. Nonrelativistic systems with different power law potentials have been considered. The results from the ET and the IET have been compared not with very accurate numerical calculations but with the HOB0 approximation described in \cite{sema20}, that is to say a variational method based on a trial three-body state built with two Gaussian functions with adjustable sizes. We know from our previous results that this method can give very satisfactory upper bounds. It appears that the mixing of attractive and negative interactions is not the cause of the poor results but instead the presence of singular potentials. In this last situation, the binding energy computed by the ET is then always of bad quality and the IET seems to make up the situation only for identical particles \cite{sema15a,sema15b}. Supplementary studies are necessary to check the relevance of these first results. 

\section{concluding remarks and prospects}
\label{sec:conclu}

The envelope theory (ET) is a remarkably easy to use method to obtain approximate but reliable eigensolutions of some quantum many-body systems \cite{hall80,hall83,hall04,silv10,silv12,sema13,sema20,sema17,sema18b,sema19}. It is particularly useful when analytical or numerical results with only a reasonable accuracy are asked for \cite{sema21}, as simple tests for numerical codes \cite{sema15a,char15}, or when a systematic study as a function of the number of particles is necessary as in the Large-$N$ approach of QCD \cite{buis11,buis12}. When all particles are identical for $D\ge 2$, the non-physical strong degeneracy inherent to this method can be partly removed by combining it with the dominantly orbital state method leading to what we called the improved envelope theory (IET). This IET can then provide improved eigenvalues \cite{sema15b}, with large effects in some cases. That is the reason why this IET is generalised in this paper to systems with identical particles plus a different one to open new domains of applicability of this method. 

The new equations to solve are explicitly written for these systems, for general kinetic parts and two-body central potentials. They have more complex structures than the ones for all identical particles, which does not allow to find analytical solutions, but they are not more complicated to solve numerically. The procedure is quite straightforward and can be easily implemented. Tests of accuracy are performed on three different systems. The main finding is that the accuracy of the eigenvalues is always improved, but the variational character can be lost. When the ET solution is very poor, the IET cannot provide an enough large improvement leading to a reasonable accuracy in some cases. Nevertheless, it seems that results obtained here already show that the IET can be safely used for a large variety of systems. 

The IET relies on the generalisation to $N$-body systems of the effective quantum number $q$ (\ref{qnq}) which gives good spectra for 2-body systems \cite{loba09}. The results obtained here show that this idea works well, at least for some Hamiltonians. But this is not a rigorous demonstration. As the determination of $q$ is performed within the WKB approximation, the same kind of study could be attempted for $N$-body systems. WKB computations are possible for such systems \cite{azbe90,niss10}. A promising idea could be to perform an hyperradial approximation of the considered Hamiltonian as in \cite{timo12,timo17} and solve it with the WKB approach. In this way, it could be possible to obtain analytical information about the quality of an effective quantum number for $N$-body systems, and possibly improve the ET in another way.

It is mentioned in Sec.~\ref{sec:intro}, that the ET and the IET can be used for various systems. A special type of three-body forces is used in clusters of cold atoms to produce a strong short-range repulsion \cite{gatt11,gatt13,kiev20} or in hadrons to take into account peculiarities of the colour interaction between quarks \cite{ferr95,dmit01,pepi02}. As this kind of three-body force can be handled by the ET \cite{sema18b}, it could be interesting to try an extension within the IET. 

The ET can give estimations of critical coupling constants in nonrelativistic $N$-body systems \cite{sema13}. Let us consider a set of identical particles with a mass $m$ interacting via a two-body potential $V(r)=-g\ v(r)$ admitting only a finite number of bound states, where $g$ is a positive quantity which has the dimension of an energy and $v(x)$ a globally positive dimensionless function vanishing at infinity. The approximate value of $g$ allowing for a bound state with quantum numbers $\{n_i,l_i \}$ is given by \cite{sema13}
\begin{equation}
\label{gcrit}
g=\frac{1}{u^2\,v(u)}\frac{2}{N(N-1)^2}\frac{Q(N)^2}{m},
\end{equation}
where $Q(N)$ is given by (\ref{QN}) and where the number $u$ is the solution of $2\, v(u)+u\, v'(u)=0$. It is then possible to link the values of the critical coupling constants between two sets with $N$ and $N+1$ particles. For a bosonic ground state (\ref{QphiBGS}), the universal relation is 
\begin{equation}
\label{gcritB}
g(N+1)=\frac{N}{N+1}g(N).  
\end{equation}
For a fermionic ground state with a large number of particles (\ref{QFGSapp}), the corresponding relation is
\begin{equation}
\label{gcritF}
g(N+1)=\left(\frac{N}{N+1}\right)^{(D-2)/D}g(N).
\end{equation}
These results can be connected with the Efimov physics \cite{naid17}. At the origin, the Efimov effect is the existence of three-body bound states whereas the interaction between the particles cannot bind two of them. But, it can be extended to a larger number of particles, and could then be studied by the ET and the IET for sets of identical or different particles. It is worth mentioning that both methods could have a very limited interest for a small number of particles for the Efimov physics. Indeed, the ET cannot find the bound state of a few identical bosons interacting via a weak Gaussian potential both at $D=1$ \cite{sema19} and $D=3$ \cite{sema15a}, whereas the bound states can be computed with a good accuracy for a large number of particles.

As a next step, we plan to perform a systematic study of the accuracy of the ET and the IET for 3-body systems containing bosons and fermions with an expansion of the eigenstates in large harmonic oscillator basis \cite{nunb77,silv96,silv00}. This will allow for a careful analysis of spectra as the one performed in \cite{sema15a} to better precise the validity of the ET and the IET. As already mentioned in \cite{sema20}, it would be desirable to have available accurate computations of eigenvalues for large many-body systems containing different particles, bosons or fermions, as the ones performed in \cite{horn14} for identical bosons, to test the ET and the IET on a larger scale.

\begin{acknowledgments}
C.C. and L.C. would like to thank the Fonds de la Recherche Scientifique - FNRS for the financial support. This work was also supported under Grant Number 4.45.10.08. All authors would like to thank Jean-Marc Richard for providing the accurate results in Table~\ref{tab:URoh} and the anonymous referee for the useful suggestions. 
\end{acknowledgments} 

\appendix

\section{Two coupled oscillators}
\label{sec:tco}
Let us consider the Hamiltonian for two one-dimensional coupled oscillators
\begin{equation}
\label{Htodco}
H=\frac{1}{2} \left( \frac{1}{\mu_a} p_1^2 + \frac{1}{\mu_b} p_2^2 + k_a\, x_1^2 + k_b\, x_2^2+ k_c\, x_1\, x_2\right),
\end{equation}
where $p_j = -i\partial/\partial x_j$ with $j=1,2$. Using the procedure described in \cite{maka18}, one can find
\begin{subequations}
\label{todco}
\begin{equation}
\label{Etodco}
E_{nn'} = \sqrt{\frac{A}{\mu}} \left(n + \frac{1}{2}\right)+ \sqrt{\frac{B}{\mu}} \left(n' + \frac{1}{2}\right), \\
\end{equation}
where
\begin{align}
\label{Etodco2}
& \mu=\sqrt{\mu_a\, \mu_b}, \nonumber \\
& \begin{cases}
A= \sqrt{\frac{\mu_b}{\mu_a}}k_a \\
B= \sqrt{\frac{\mu_a}{\mu_b}}k_b
\end{cases} \quad \text{if}\quad  k_c = 0, \nonumber\\
& \begin{cases}
A= \sqrt{\frac{\mu_b}{\mu_a}}k_a - \frac{k_c}{2} \\
B= \sqrt{\frac{\mu_b}{\mu_a}}k_a + \frac{k_c}{2}
\end{cases} \quad  \text{if} \quad \epsilon= \frac{1}{k_c}\left(\sqrt{\frac{\mu_a}{\mu_b}}k_b - \sqrt{\frac{\mu_b}{\mu_a}} k_a \right)= 0, \\
& \begin{cases}
A= \sqrt{\frac{\mu_b}{\mu_a}}k_a - \frac{k_c}{2}\left(\textrm{sgn}(\epsilon)\sqrt{1+\epsilon^2}-\epsilon \right) \\
B= \sqrt{\frac{\mu_a}{\mu_b}}k_b + \frac{k_c}{2}\left(\textrm{sgn}(\epsilon)\sqrt{1+\epsilon^2}-\epsilon \right)
\end{cases} \quad \text{if} \quad \epsilon \neq 0, \nonumber
\end{align}
\end{subequations}
with
\begin{equation}
\label{limeps}
\lim_{\epsilon\,\to\, 0^{\pm}} \left(\textrm{sgn}(\epsilon)\sqrt{1+\epsilon^2}-\epsilon \right) = \pm 1.
\end{equation}
Let us note that the equivalent formulas given in \cite{maka18} are only valid if $\mu_a = \mu_b$. 

\section{Bosonic and fermionic Ground states}
\label{sec:gs}

An approximate eigenstate obtained with the ET is an eigenstate of a many-body harmonic oscillator Hamiltonian. So the symmetry and other characteristics of the approximate solutions are given by the properties of the solutions of this Hamiltonian. If we consider only the spatial symmetry of the many-body states, it is shown that the internal energy of a system of $N$ identical oscillators, interacting via two-body harmonic forces, is equivalent to the energy of a system of decoupled oscillators in a common central field, provided the contribution of the centre of mass is removed \cite{levy68}. Using this result, the bosonic ground state (BGS) for the ET solution for $N$ identical particles is trivially given by 
\begin{equation}
\label{QBGS}
Q_\textrm{BGS}(N) = (N-1)\frac{D}{2},
\end{equation}
since all quantum numbers are vanishing. For fermions with degeneracy $d$, the situation is more complicated. In this case, the fermionic ground state (FGS) can be computed with the two following relations   
\begin{subequations}
\label{QFGS}
\begin{align}
\label{QFGS1}
Q_\textrm{FGS}(N) &= d\, D \binom{q + D-1}{D+1} + q\, r + (N-1)\frac{D}{2} \quad \textrm{with} \\
\label{QFGS2}
N &= d \binom{q + D-1}{D} + r,
\end{align}
\end{subequations}
where $q$ is the greatest natural number such that $r\ge 0$ in (\ref{QFGS2}) \cite{sema20}. 

How to modify these formulas when the number $\phi$ is introduced in the global quantum number $Q$? We consider that the energy is still obtained using the rules defined in \cite{levy68}. But in piling the particles on levels with higher and higher energies, we must take into account that the presence of $\phi$ modifies the order of the levels with respect to the usual harmonic oscillator levels. For bosons, the problem is still trivial since all quantum numbers can be put to zero
\begin{equation}
\label{QphiBGS}
Q_{\phi\,\textrm{BGS}}(N) = (N-1)\frac{D+\phi-2}{2}.
\end{equation}
For fermions, when $\phi=1$, it is possible to perform a calculation similar to the one giving (\ref{QFGS}). The first step is to determine the degeneracy $W_{l}^D$ of a level of orbital angular momentum $l$ in $D$ dimensions ($D\ge 2$) \cite{yane94} 
\begin{equation}
W_{l}^D =\sum_{m_1=0}^{l} \sum_{m_2=0}^{m_1}... \sum_{m_{D-2}=-m_{D-3}}^{m_{D-3}} d = d\ \frac{2l+D-2}{D-2} \binom{l+D-3}{D-3},
\end{equation}
where the $\{m_i\}$ are the magnetic quantum numbers. This formula can be demonstrated by induction. One can check that $W_l^2=d(2-\delta_{l0})$ and that $W_l^3=d(2l+1)$, as expected. Then, the fermionic ground state $Q_{\phi\,\textrm{FGS}}(N)$ for $\phi=1$ can be computed with the two following relations   
\begin{subequations}
\label{QphiFGS}
\begin{align}
\label{QphiFGS1}
Q_{1\,\textrm{FGS}}(N) &= d \ \frac{(2\,q\, D -2\, D+D^2+1)}{D+1}  \binom{q+D-2}{D} + q\, r + (N-1)\frac{D-1}{2}
\quad \textrm{with} \\
\label{QphiFGS2}
N &= d\ \frac{2\, q+D-2}{D} \binom{q+D-2}{D-1} + r,
\end{align}
\end{subequations}
where $q$ is the greatest natural number such that $r\ge 0$ in (\ref{QphiFGS2}). 

When $\phi$ is arbitrary, we must resort to a numerical routine to determine $Q_{\phi\,\textrm{FGS}}(N)$. We have checked that this routine is in agreement with formulas~(\ref{QFGS}) and (\ref{QphiFGS}). We have noticed that the following approximate formula
\begin{equation}
\label{QFGSapp} 
Q_{\phi\,\textrm{FGS}}(N) \approx \frac{D}{D+1}\left(\frac{\phi\,D!}{2\,d}\right)^{1/D}N^{(D+1)/D}
\end{equation}
is valid for small values of $D$, $d$ and $\phi$ when $N \gg 1$. Note that the bosonic and fermionic ground states are the same if $d\ge N$.

\end{document}